# A Potential Skyrmion Host Fe(IO$_3$)$_3$: Connecting Stereo-active Lone-Pair Electron Effects to Dzyaloshinskii-Moriya Interaction

Ebube E. Oyeka,[1] Michał J. Winiarski,[2] Artur Błachowski,[3] Keith M. Taddei,[4] Allen Scheie,[4] and Thao T. Tran*[1]

[1]Department of Chemistry, Clemson University, Clemson, SC 29634, USA

[2]Faculty of Applied Physics and Mathematics and Advanced Materials Center, Gdansk University of Technology, ul. Narutowicza 11/12, 80-233 Gdansk, Poland

[3]Mössbauer Spectroscopy Laboratory, Institute of Physics, Pedagogical University of Cracow, ul. Podchorążych 2, 30-084 Kraków, Poland

[4]Neutron Scattering Division, Oak Ridge National Laboratory, 9500 Spallation Dr, Oak Ridge, TN 37830, USA

**ABSTRACT:** Magnetic skyrmions, which are topologically distinct magnetic spin textures, are gaining increased attention for their unique physical properties and potential applications in spintronic devices. Here we present a design strategy for skyrmion host candidates based on combinations of magnetic spin, asymmetric building units having stereo-active lone-pair electrons, and polar lattice symmetry. To demonstrate the viability of the proposed rational design principles, we successfully synthesized Fe(IO$_3$)$_3$ polycrystalline sample and single crystals by using a new simplified low-temperature pathway, which is experimentally feasible for extending materials growth of transition metal iodates. Single crystal X-ray and powder synchrotron X-ray diffraction measurements demonstrated that Fe(IO$_3$)$_3$ crystallizes in the polar chiral hexagonal lattice with space group $P6_3$. The combined structural features of the macroscopic electric polarization along the $c$-axis stemming from the co-alignment of the stereo-active lone-pairs of the IO$_3^-$ trigonal pyramid and the magnetic Fe$^{3+}$ cation residing on the three-fold rotation axis were selected to promote asymmetric exchange coupling. We find evidence of a predicted skyrmion phase at 14 K ≤ $T$ ≤ 16 K and 2.5 T ≤ $\mu_0 H$ ≤ 3.2 T driven by Dzyaloshinskii–Moriya (DM) interaction, a conclusion supported by the appreciable DM exchange and the zero-field spiral antiferromagnetic ground state of Fe(IO$_3$)$_3$ deduced from neutron diffraction experiments. The associated magnetic modulation wavelength of the putative skyrmions is expected to be short ~ 18 nm, comparable to the period of the DM-driven incommensurate order. This work links stereo-active lone-pair electron effects to enhanced DM interaction, demonstrating a new approach for chemical guidelines in the search for skyrmionic states of matter.

## INTRODUCTION

Magnetic skyrmions, which are spatially inhomogeneous spin textures with a nano-sized particle character, are an exciting platform for new spin topology physics. Their unique stability and nano-size are promising not just for the study of new phenomena intrinsically related to topological states of matter, but also for spin-based technologies wherein information is processed and transmitted by the electron spin rather than the electron charge.[1-3] The topological property of skyrmions allows them to be not just efficiently controlled by various external stimuli such as electric currents in metals and electric fields in insulators but also robust with defects or lattice imperfections.[1-3]

Despite deep interest from the physics and materials chemistry communities, only a limited number of materials hosting skyrmions have been realized, including, B20-structure (MnSi, FeGe, Fe-Co-Si), $\beta$-Mn-type (Co-Zn-Mn, FePd$_{1-x}$Pt$_x$Mo$_3$N), GaV$_4$S$_8$, GaV$_4$Se$_8$, Cu$_2$OSeO$_3$ and VOSe$_2$O$_5$.[4-15] B20 metal compounds such as MnSi, FeGe, Fe-Co-Si (space group $P2_13$) were first observed to feature 3D skyrmion lattice characterized by emergent magnetic monopoles.[5, 6, 8, 16] $\beta$-Mn-type materials including Co-Zn-Mn, FePd$_{1-x}$Pt$_x$Mo$_3$N, which crystallize in space group $P4_132/P4_332$ – slightly higher, symmetry than for B20 magnets yet also chiral cubic, exhibit skyrmion phases at and above room temperature.[7, 9, 10] In addition to these metals, magnetic skyrmions can also be realized in insulating materials. GaV$_4$S$_8$, a small-gap semiconductor having polar trigonal symmetry $R3m$, possesses radial spin textures.[17, 18] A combined experimental and computational study of GaMo$_4$Se$_8$ showed that the ground state $R3m$ phase of this material hosts Neel-type skyrmions at 2 K < $T$ < 28 K while a metastable $Imm2$ phase features uniaxial ferromagnetism.[19] In the insulating chiral magnet Cu$_2$OSeO$_3$ ($P2_13$),[11] skyrmions induce electric polarization and can be controlled by external electric fields.[3]

In these crystal structure types, breaking inversion center symmetry promotes asymmetric Dzyaloshinskii–Moriya (DM) exchange interaction $H_{DM} = D \cdot (S_i \times S_j)$ facilitated by spin-orbit coupling (SOC).[20, 21] The DM interaction stabilizes different forms of non-trivial magnetic spin textures depending on the underlying lattice symmetry.[22, 23] Competing interactions between this DM exchange and the Heisenberg magnetic interaction can stabilize topologically protected spin states.[24]

Designing and creating materials featuring skyrmions, however, presents a significant challenge. This is in part due to the limited ability to simultaneously control both local and extended structure, as well as the undesired electron filling of the frontier orbitals.[25] Here, we show that these problems can be overcome by using judicious chemical design of metal ion and anion group with stereo-active lone-pair electrons, that enables tunability of DM exchange interaction thereby potentially facilitating evolution of such novel spin states.

## EXPERIMENTAL SECTION

**Reagents.** $FeCl_3.6H_2O$ (Alfa Aesar, 99.5 %), $HIO_3$ (Alfa Aesar, 99.5 %), $HNO_3$ (Alfa Aesar, 67 %), $GaCl_3$ (Alfa Aesar, 99.99 %) were used as starting materials.

**Synthesis.** Polycrystalline of $Fe(IO_3)_3$ was synthesized by heating amorphous $Fe(IO_3)_3.6H_2O$ at 380 °C for 8 hours. For amorphous $Fe(IO_3)_3.6H_2O$, $FeCl_3.6H_2O$ (2.5 mmol) was dissolved in $HNO_3$ (1M, 25 mL) and then an aqueous solution of $HIO_3$ (7.5 mmol, 10 mL) was added. The mixture was stirred at 80 °C for 1 h. Amorphous solid $Fe(IO_3)_3.6H_2O$ was isolated from the mother-liquor by filtration and washed with deionized water (Yield of $Fe(IO_3)_3$, 89 %; based on Fe). Single crystals of $Fe(IO_3)_3$ were synthesized by hydrothermal method. $FeCl_3.6H_2O$ (2.5 mmol), $HIO_3$ (7.5 mmol) and $HNO_3$ (1M, 10 mL) were placed in a 23 mL-Teflon-lined autoclave. The autoclave was heated at 200 °C for 120 h and cooled slowly to 25 °C at a rate of 5 °C/h. Yellowish green rod-shaped crystals of $Fe(IO_3)_3$ were isolated by filtration and washed with deionized water (Yield of $Fe(IO_3)_3$, 82 %; based on Fe).

Polycrystalline of $Ga(IO_3)_3$ was synthesized by heating amorphous $Ga(IO_3)_3.13H_2O$ at 400 °C for 4 hours. For amorphous $Ga(IO_3)_3.13H_2O$, $GaCl_3$ (2.5 mmol) was dissolved in $HNO_3$ (1M, 25 mL) and then an aqueous solution of $HIO_3$ (7.5 mmol, 10 mL) was added. The mixture was stirred at 80 °C for 1 h. Amorphous solid $Ga(IO_3)_3.13H_2O$ precipitated after one week as colorless solid and was isolated from the mother-liquor by filtration and washed with deionized water. Anhydrous polycrystalline $Ga(IO_3)_3$ has pale orange color (Yield of $Ga(IO_3)_3$, 56 %; based on Ga).

**Synchrotron X-ray diffraction.** Synchrotron XRD pattern of $Fe(IO_3)_3$ was collected using the 11-BM beamline at Advanced Photon Source, Argonne National Laboratory. Data were collected at $T$ = 295 K and $\lambda$ = 0.45789 Å. No impurities were observed. Rietveld refinement of XRD pattern was performed using TOPAS Academic V6. Vesta software was used for crystal structure visualization (Figure S1).[26]

**Powder X-ray diffraction.** Powder X-ray diffraction (PXRD) measurements on $Ga(IO_3)_3$ was performed using Rigaku Ultima IV diffractometer equipped with Cu $K_\alpha$ radiation ($\lambda$ = 1.5406 Å). Data were collected in the 2$\theta$ range of 5° – 90° at 0.2°/min. Rietveld refinement of XRD pattern was performed using TOPAS Academic V6.

**Single crystal X-ray diffraction.** Single crystal diffraction experiments were performed on $Fe(IO_3)_3$ using a Bruker D8 Venture diffractometer with Mo $K_\alpha$ radiation ($\lambda$ = 0.71073 Å) and a Photon 100 detector at $T$ = 100 and 300 K. Data processing (SAINT) and scaling (SADABS) were performed using Apex3 software system. The structure was solved by intrinsic phasing (SHELXT) and refined by full matrix least-squares techniques on F2 (SHELXL) using the SHELXTL software suite.[27] All atoms were refined anisotropically.

**Thermal analysis.** Thermogravimetric analysis and differential scanning calorimetry measurements were performed using a TA SDT Q600 Instrument. Approximately 10 mg of $Fe(IO_3)_3.6H_2O$ was placed in an alumina crucible and heated at a rate of 20 °C/min from room temperature to 1000 °C under flowing nitrogen (flow rate: 100 mL/min) (Figure S2).

**Infrared spectroscopy.** Attenuated total reflection Fourier transform infrared (ATR-FTIR) spectrum for $Fe(IO_3)_3$ was collected on a Shimadzu IR Affinity-1S in the range of 400 to 4000 cm$^{-1}$ (Figure S3).

**Magnetization and specific heat.** DC magnetization measurements on $Fe(IO_3)_3$ powder were performed with the Vibrating sample magnetometer (VSM) option of Quantum Design Physical Properties Measurement System (PPMS). AC magnetization was measured using the ACMS option (Figure S4). Magnetic susceptibility was approximated as magnetization divided by the applied magnetic field: $\chi \approx M/H$. Heat capacity was measured using the PPMS, employing the semiadiabatic pulse technique from $T$ = 2 K to 300 K (Figure S5).

**Neutron Diffraction.** Neutron diffraction was measured on a 5 g loose powder $Fe(IO_3)_3$ sample on the HB2A powder diffractometer at ORNL's HFIR reactor.[28] The sample was inside an aluminum can and mounted on an orange cryostat with a sample rotator stick. Data were measured at $T$ = 1.6 K, 20 K, and 150 K for two hours each. Magnetic refinements were performed both with and without the data at $T$ = 150 K subtracted, and the results were the same to within error bars, but the high-temperature subtraction allowed for a clearer visualization of the fit and a lower overall $\chi^2$. To perform this temperature subtraction, some of the strong high-angle 150 K nuclear Bragg peaks were shifted in $Q$ so that the high and low temperature nuclear peaks peaked at exactly the same scattering vector (thermal expansion slightly shifted the 150 K Bragg peaks relative to 1.6 K).

To identify the magnetic propagation vector, we made plots of the expected magnetic Bragg peaks as a function of propagation vector, as shown in Figure S6 of Supporting Information. This allows us to visually identify the correct propagation vector. The only magnetic propagation vector which indexes all observed peaks without adding extra peaks is $q = (0, 0, 0.0293(1))$. Using this propagation vector, we performed an irreducible representation decomposition as explained in the main text. The fits to magnetic structure are plotted in Figure S7-S8, and the $\Gamma_2 / \Gamma_3$ structure clearly matches the data the best.

As shown by the 20 K scattering in Figure S9, there are measurable magnetic correlations even above the Neel temperature. To estimate the correlation length, we convolved the 1.6 K data with a Lorentzian to model short-ranged magnetic correlations. We found that a correlation length of 3.3 Å matches the 20 K data very well, as shown in Figure S9b. Thus, the low-temperature paramagnetic state has the same magnetic structure as the long range ordered phase, but with a very short correlation length (shorter, in fact than the nearest neighbor Fe-Fe distance).

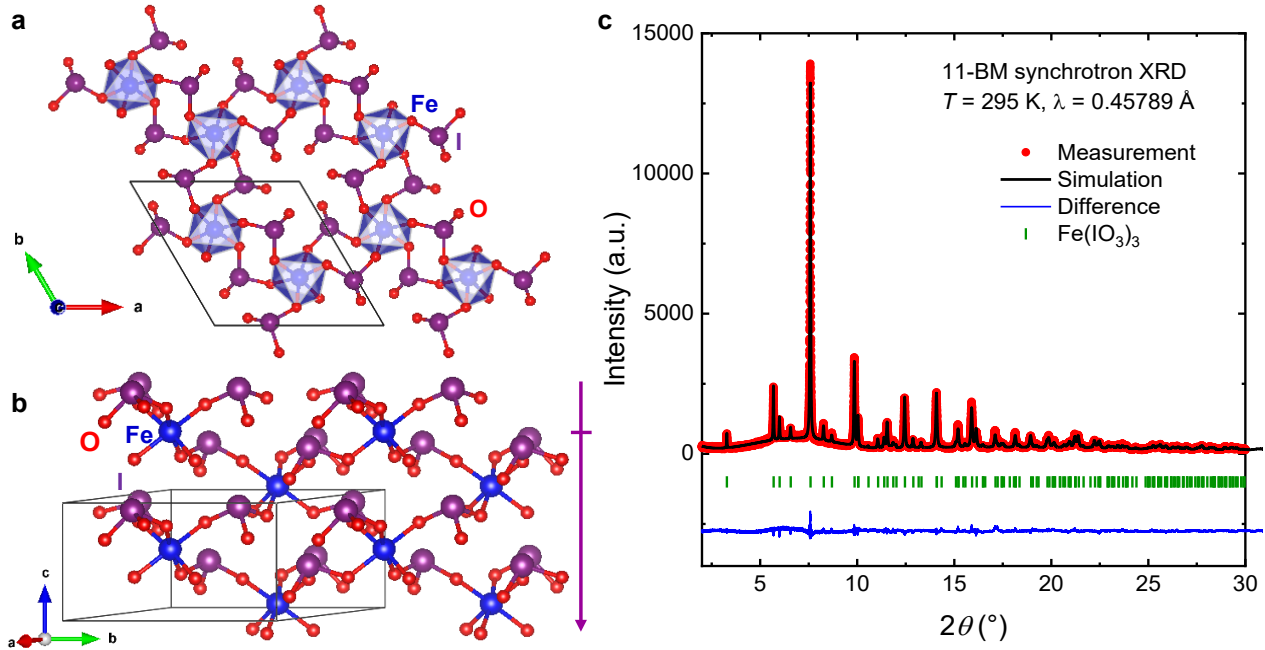

Figure 1. (a-b) Crystal structure of $Fe(IO_3)_3$ consisting of corner-shared $FeO_6$ octahedra and $IO_3$ trigonal pyramid; The local dipole moment of $IO_3$ is additive and results in a macroscopic electric polarization along the *c*-axis direction; (c) Rietveld refinements of 11-BM synchrotron XRD data of polycrystalline $Fe(IO_3)_3$.

**$^{57}$Fe Mössbauer spectroscopy.** $^{57}$Fe Mössbauer spectroscopy measurements were performed in transmission mode using the RENON MsAa-4 spectrometer equipped with the LND Kr-filled proportional detector. A commercial $^{57}$Co(Rh) source made by RITVERC GmbH was applied. The source line-width $\Gamma_s$ = 0.106(5) mm/s and the effective source recoilless fraction were derived from the fit of the Mössbauer spectrum of the 10-μm-thick α-Fe foil. The Mössbauer absorber was prepared using 30 mg of the studied sample mixed with the $B_4C$ carrier and the absorber thickness (surface density) amounted to 15 mg/cm$^2$ of investigated material. The Janis Research Inc. SVT-400 cryostat was used to maintain temperature of absorber in the range 5 K – 300 K, with the long-time accuracy better than 0.1 K (except for 5 K, where the standard deviation was about 0.5 K). Obtained data were processed by means of the MOSGRAF software suite using the transmission integral approximation.

**DFT calculation.** Electronic structure calculations were performed by means of the density functional theory (DFT) using the Quantum Espresso[29] software package employing the Generalized Gradient Approximation (GGA) of the exchange-correlation potential with the PBEsol parametrization.[30] Scalar-relativistic projector-augmented wave (PAW) potentials[31] for Fe, I, and O were taken from the PSlibrary v. 1.0.0 set.[32] A 5x5x9 Monkhorst-Pack *k*-point mesh was used for calculations. Kinetic energy cutoff for charge density and wavefunctions was set to 90 eV and 1080 eV, respectively. A 4.0 eV on-site repulsion term *U* was applied to *d*-shells of Fe atoms.[32]

## RESULTS AND DISCUSSION

$Fe(IO_3)_3$ was synthesized by combining low-temperature chemistry technique (*T* = 200 °C for single crystals; *T* = 380 °C for polycrystalline form) and thermal analysis (Figure S1-S2). Compared to the previously reported synthesis of $Fe(IO_3)_3$ single crystals under extreme conditions of high pressure and high temperature (*P* = 500 $O_2$ atm, *T* = 520 °C),[33] our new method provides a more facile pathway to prepare high-quality bulk sample and single crystalline forms of $Fe(IO_3)_3$, potentially leading to the creation of many new transition-metal iodates.

$Fe(IO_3)_3$ crystal structure consists of corner-sharing $FeO_6$ octahedra and $(IO_3)^-$ trigonal pyramids (Figure 1). Each $Fe^{3+}$ (*S* = 5/2, 3$d^5$) cation is bonded to six oxygen atoms in a slightly trigonal distortion of octahedral coordination with Fe–O distances of 2.055(3) Å and 2.006(2) Å. Each $I^{5+}$ cation is coordinated with three oxygen atoms in trigonal pyramidal geometry, attributed to the stereo-active lone-pair electrons of $I^{5+}$.[34] The iodate I–O bond distances range from 1.798(2) Å to 1.921(3) Å. The subtle trigonal distortion of the FeO$_6$ octahedron is likely induced by the stereo-active lone-pair electrons of the I$^{5+}$ cation.[34-36] The polarity of the chiral hexagonal structure of Fe(IO$_3$)$_3$ (space group $P6_3$) is driven by the uniform alignment of the trigonal pyramid (IO$_3$)$^-$ units (Figure 1a-b). The crystal structure obtained from Rietveld refinements of high-resolution X-ray synchrotron (11-BM) powder diffraction data is in excellent agreement with that obtained from single crystal X-ray diffraction (Figure 1c and Table S1).

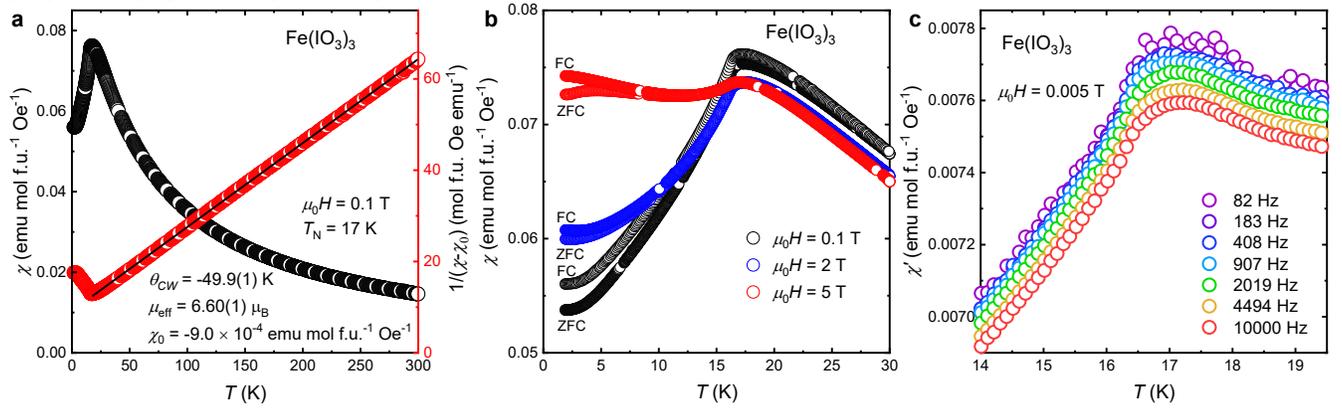

Figure 2. (a) (black) DC magnetic susceptibility of Fe(IO$_3$)$_3$ and (red) Curie-Weiss fit of $1/(\chi-\chi_0)$ against temperature for the paramagnetic phase of Fe(IO$_3$)$_3$ at $\mu_0H$= 0.1 T. (b) Zero-field-cooled (ZFC) and field-cooled (FC) magnetic susceptibilities of Fe(IO$_3$)$_3$ as a function of temperature at different magnetic field. (c) Frequency dependence of AC susceptibility as a function of temperature.

The occurrence and alignment of the lone-pair strongly influences the crystal symmetry and related properties of Fe(IO$_3$)$_3$. The polarization attributable to the Fe$^{3+}$ octahedra is negligible as the distortion is very minute. With the (IO$_3$)$^-$ trigonal pyramid, however, the local dipole moment is additive and results in a macroscopic electric polarization along the c-axis direction as seen in Figure 1b (and Figure S1).

The local structure symmetry of the asymmetric building unit of (IO$_3$)$^-$ is confirmed to be $C_{3v}$, $\Gamma_{vib} = 2A_1 + 2E$, as evidenced by the four fundamental I–O vibrational modes observed at 810, 756, 683 and 551 cm$^{-1}$ in the FTIR spectrum (Figure S3). In addition, iodine is a heavy element and exhibits increased SOC that can influence electronic properties through splitting of degenerate orbitals, producing unusual complex effects on magnetic ordering vectors in materials. Thus, utilizing and controlling the packing of the (IO$_3$)$^-$ asymmetric building units provides a strategy for breaking inversion symmetry in the crystal lattice of magnetic materials.

The temperature dependence of the magnetic susceptibility of Fe(IO$_3$)$_3$ shows unusual behavior (Figure 2). The material undergoes a transition to an antiferromagnetic (AFM) order at $T_N$ = 17 K (Figure 2a). $T_N$ changes little with applied magnetic fields 0.1 T < $\mu_0H$ < 5 T, and zero-field-cooled (ZFC) and field-cooled (FC) susceptibilities show a bifurcation at $T < T_N$. Under $\mu_0H$ = 5 T, Fe(IO$_3$)$_3$ possesses magnetization upturn at $T \sim 12$ K (Figure 2b), that is likely attributable to the presence of competing AFM-FM interactions potentially giving rise to metamagnetic transition, i.e., a transition from the AFM state to a spin-polarized state. AC magnetization results show that the transition temperature does not change when the excitation frequency is varied from 82 Hz to 10 kHz (Figure 2c). This confirms the AFM character of the transition at low magnetic fields, excluding the possibility of spin-glass-like behavior. At $T > 17$ K, the inverse magnetic susceptibility is well described by the Curie-Weiss law for paramagnetic spins (Figure 2a).[37] The Curie-Weiss temperature $\theta_{CW}$ = -49.9(1) K was extracted from the intercept of the linear fit. The observed transition temperature $T_N$ = 17 K is almost 3 times lower than the Curie-Weiss temperature $\theta_{CW}$, indicating significant magnetic frustration (presence of competing exchange interactions) between Fe$^{3+}$ magnetic moments. The effective magnetic moment $\mu_{eff}$ of the Fe$^{3+}$ cation estimated from the resulting Curie constant is 6.60(1) $\mu_B$, which is slightly increased from the ideal $g(S(S+1))^{1/2}$ = 5.92 $\mu_B$ value expected for a free $S$ = 5/2 moment. This may be due to a partial orbital contribution to the moment from SOC. The metamagnetic behavior of Fe(IO$_3$)$_3$ is evident by the hysteresis loop of the $M(H)$ curve (Figure S4), indicating a transition from AFM order to field-polarized magnetism.[38]

The formation of skyrmions is typically associated with a positive entropy change since magnetic skyrmions form as a high entropy state. The dc magnetization technique is applied for studying the magnetoentropic signatures of skyrmion materials,[39] demonstrating that the field-induced entropic response can be understood through a skyrmion phase and a boundary of first-order phase transitions expected for topologically non-trivial spin states.[16, 39] The isothermal entropy change upon magnetization $\Delta S_M(H,T)$ is obtained from the Maxwell relation:

$$(dM/dT)_H = (dS/dH)_T \quad (1)$$

where $S$ is total entropy, $H$ is magnetic field, $M$ is magnetization, and $T$ is temperature.

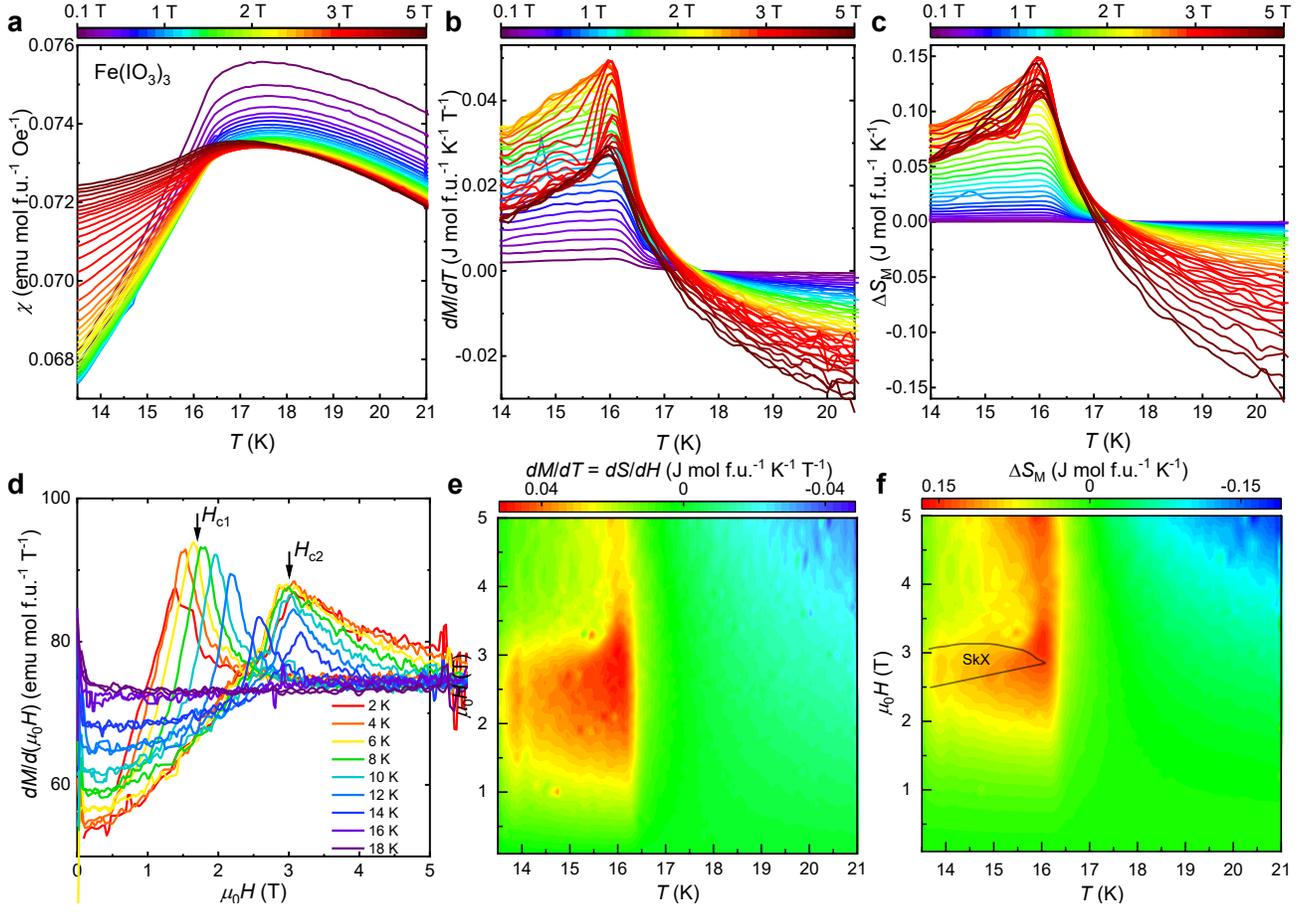

Figure 3. (a) χ(T) (M(T) / (μ₀H)) curves of Fe(IO$_3$)$_3$ at different magnetic field. (b) First derivative of magnetization with respect to the temperature, $dM/dT$, for Fe(IO$_3$)$_3$. (c) Isothermal magnetic entropy of Fe(IO$_3$)$_3$ at each different magnetic field obtained by the integral of $dM/dT$ with respect to the magnetic field. (d) First derivative of magnetization with respect to the magnetic field $dM/dH$ curve of Fe(IO$_3$)$_3$. (e-f) Magnetoentropic map of Fe(IO$_3$)$_3$ near the ordering temperature. (e) A map of $dM/dT = dS/dH$ for Fe(IO$_3$)$_3$ reveals clear ridges (red) and valleys (green) indicating first-order transitions. (f) A map of $\Delta S_M(T,H)$ of Fe(IO$_3$)$_3$. The boundary of the skyrmion phase was extrapolated from the critical magnetic fields, $H_{c1}$ and $H_{c2}$ obtained from the $dM/dH$ curves.

The magnetoentropic mapping of Fe(IO$_3$)$_3$ reveals the emergence of a putative skyrmion phase observed at 14 K ≤ T ≤ 16 K and 2.5 T ≤ μ₀H ≤ 3.2 T, highlighting the feasibility of the design strategy for inducing such topologically distinct spin states in new classes of materials.

Figure 3a illustrates how the magnetization evolves as a function of temperature under a series of applied fields near the magnetic ordering temperature. The rich magnetic behavior of Fe(IO$_3$)$_3$ is depicted in the first derivative of magnetization with respect to temperature $dM/dT$ at each magnetic field (Figure 3b) and the final integrated $\Delta S_M$ curves (Figure 3c). The magnetic anomalies in the $dM/dT$ curves are consistent with the critical magnetic fields, $H_{c1}$ and $H_{c2}$ obtained from the $dM/dH$, indicating formation of topologically distinct magnetic spin states (Figure 3d).[16] The critical magnetic fields $H_{c1}$ and $H_{c2}$ (2.5 – 3.2 T) of Fe(IO$_3$)$_3$ are similar to those observed in MnGe and MnSc$_2$S$_4$. While these values are smaller than those of MnSi$_{1-x}$Ge$_x$ (x ≥ 3, $H_c$ = 10 – 20 T), they are larger compared to those of other skyrmion host materials.[13, 40-44] Critical magnetic fields represent an estimate of the strength of driving force for winding spins, and such diverse observation of these experimental values of $H_c$ cannot be explained by the simple model based on DM exchange alone. It could be attributed to an intricate ensemble of DM exchange, Heisenberg interaction, spin-orbital and electron-lattice coupling effects. Further study, however, is encouraged to elucidate the origin of these phenomena. Thermodynamic capacity can be elucidated in $dM/dT = dS/dH$ curves, providing complementary insights to specific heat measurements. Ridges and valleys in $dS/dH$ can imply field-driven first-order phase transitions and can provide estimates for entropy changes associated with these transitions. The $dM/dT = dS/dH$ and $\Delta S_M(H,T)$ maps of Fe(IO$_3$)$_3$ (shown in Figure 3e-f) were constructed from temperature-dependent magnetization measurements at different applied magnetic fields using the relation:

$$\Delta S_M(H,T) = \int_0^H \left(\frac{dM}{dT}\right)_{H'} dH' \qquad (2)$$

In the $dM/dT = dS/dH$ map (Figure 3e), the region at $T > T_N$ = 17 K is green transitioning to blue at higher fields, indicating a decrease in entropy which is consistent with a zero-field short-range correlated state from frustrated interactions. At $T < T_N$ = 17 K, ridges (red color) are associated with phase transitions. In the $\Delta S_M(H,T)$ map (Figure 3f), the phase diagram is represented in entropy changes corresponding to first-order transitions. The yellow nearly vertical line near $T \approx 16.5$ K depicts an onset of first-order phase transition between the magnetically ordered state and the paramagnetic state. Below this transition, there is a finite-field region of increased entropy of approximately 0.15 J mol$^{-1}$ K$^{-1}$. This bears striking resemblance to finite-field skyrmion phases seen in other materials.[45]

Furthermore, the positive entropy anomaly that starts near $T \approx 16.5$ K and broadens upon cooling shows a finite pocket of increased disorder, which is also consistent with a skyrmion phase. The evolution of a skyrmion phase is distinct from other topologically trivial states such as helical and conical phases, which can form without a change in entropy from a paramagnetic or fluctuation-disordered phase at zero-field. The putative skyrmion phase is estimated to be 14 K ≤ $T$ ≤ 16 K (2 K in width) and 2.5 T ≤ $\mu_0 H$ ≤ 3.2 T (0.7 T in height) (Figure 3f).

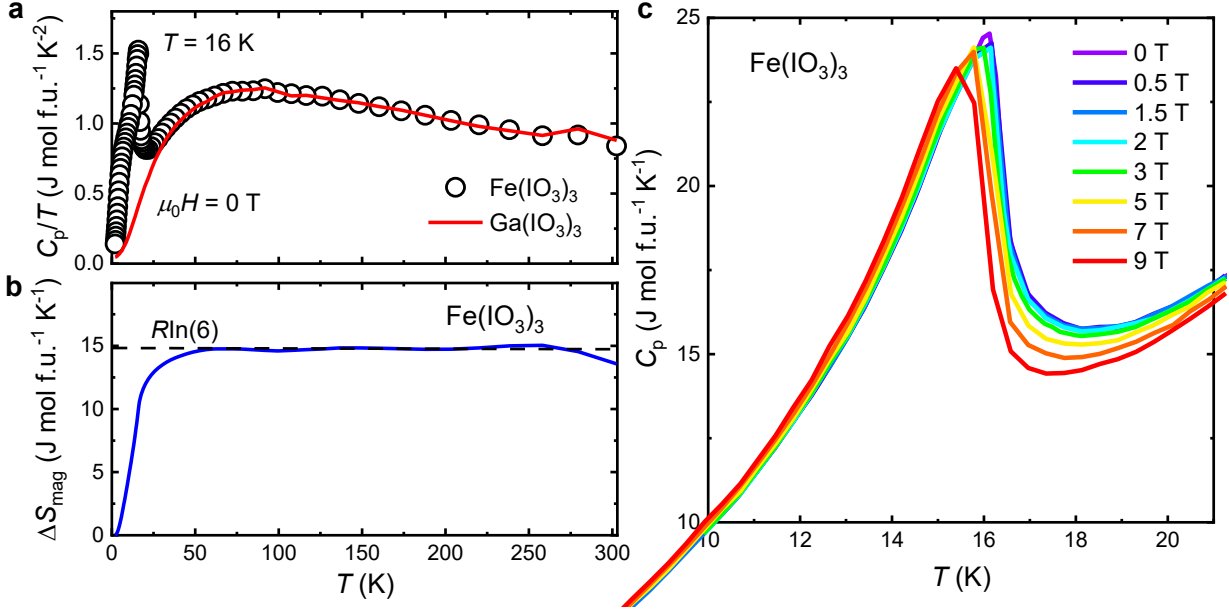

Figure 4. (a) Molar heat capacity over temperature ($C_p/T$) vs. temperature for Fe(IO$_3$)$_3$. The anomaly is consistent with the magnetic phase transition of the material. (b) Magnetic entropy change $\Delta S_{mag}$ = 14.9(1) J mol$^{-1}$ K$^{-1}$ ~ the expected value of $S = 5/2$ spins (Rln6). (c) Molar heat capacity of Fe(IO$_3$)$_3$ measured at different magnetic field $\mu_0 H$ up to 9 T. The magnetic transition temperature decreases as the applied magnetic field increases.

To elucidate the thermodynamics of the ground state of Fe(IO$_3$)$_3$, specific heat measurements were performed over the range of 2 K ≤ $T$ ≤ 300 K at zero magnetic field (Figure 4). An anomaly resembling a second-order phase transition is observed at $T \approx 16$ K in the $C_p/T$ vs. $T$ plot, which is the approximate temperature of the magnetic phase transition of the material ($T_N$ = 17 K). The entropy recovered $\Delta S$ from the transition is estimated from:

$$\Delta S = \int_0^T \frac{C_V}{T} dT \qquad (3)$$

where $C_v$ is the heat capacity at constant volume, which is approximated to be $C_p$ (heat capacity at constant pressure) for solid at low temperatures and $T$ is the temperature.[37, 46] Ga(IO$_3$)$_3$, a nonmagnetic isostructural material, was used to account for phonon contribution, allowing an accurate estimation of magnetic specific heat and entropy. The entropy change at the magnetic phase transition was estimated from the specific heat to be $\Delta S_{mag}$ = 14.9(1) J mol$^{-1}$ K$^{-1}$, in agreement with $S = 5/2$ value $R\ln(6)$ = 14.90 J mol$^{-1}$ K$^{-1}$. This indicates that the phase transition is to a magnetic ordered phase.

The magnetic phase diagram of Fe(IO$_3$)$_3$ is proposed by combining the results obtained from the magnetic susceptibility, $dM/dT$, $dM/dH$, and specific heat as shown in Figure 5.

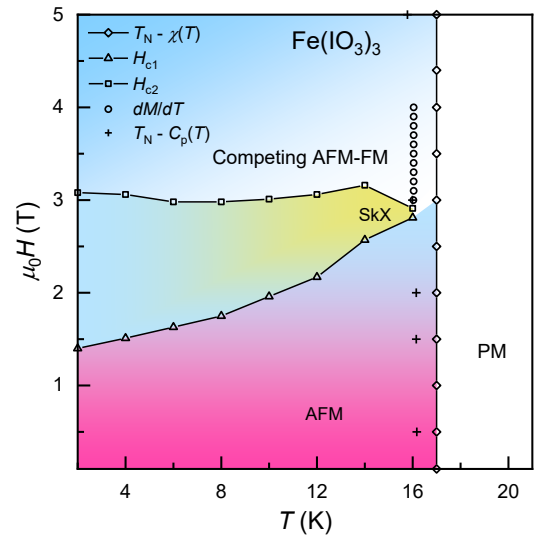

Figure 5. Magnetic phase diagram of Fe(IO$_3$)$_3$, estimated by combining magnetization, magnetoentropic mapping and specific heat measurements

In order to gain further insight into the nature of magnetic order, $^{57}$Fe Mössbauer spectroscopy studies of Fe(IO$_3$)$_3$ were performed. $^{57}$Fe Mössbauer spectra for selected temperatures are shown in Figure 6. In a paramagnetic state, the main component of spectra designated to Fe(IO$_3$)$_3$ has a shape of the narrow single line with center shift $\delta$ = 0.4 mm/s at $T$ = 300 K, typical for high-spin trivalent iron. Relatively small line-width $\Gamma$ = 0.2 mm/s and lack of measurable quadrupole splitting indicate that iron atoms occupy one crystallographic position and the nearest surrounding of the Fe$^{3+}$ ions are only subtly distorted without measurable electric field gradient. The contribution of very broad spectral component designated as Fe(IO$_3$)$_3$-$r$ (relaxation) decreases with temperature and it enhances the narrow Fe(IO$_3$)$_3$ main component. The broad spectral component of Fe(IO$_3$)$_3$-$r$ shows some manifestation of thermally dependent electronic relaxation phenomena. However, the microscopic mechanism responsible for the relaxation (e.g., spin-phonon and spin-spin interactions) or thermally controlled iron ion dynamics associated with the oscillations is difficult to predict.[47]

$^{57}$Fe Mössbauer spectra of Fe(IO$_3$)$_3$ obtained at $T$ = 12 K and 5 K show a six-line structure due to the Zeeman magnetic hyperfine splitting. The hyperfine magnetic field $B$ at 5 K with value of 51.18 T is smaller than 51.64 T obtained at 4.3 K by M. Ristić *et al.*,[48] which means that the $T$ = 5 K Fe hyperfine field varies at 0.6 TK$^{-1}$, indicating that saturation is not yet reached. A significantly lower value of the hyperfine field $B$ = 40 T at $T$ = 12 K indicates about a 30 % increase of the effective magnetic moment in the temperature range between 12 K and 5 K. A small value of the electric quadrupole splitting $\Delta$ detected at the lowest temperatures confirms that the nearest surrounding of the iron atom is slightly distorted and in a state of magnetic order.

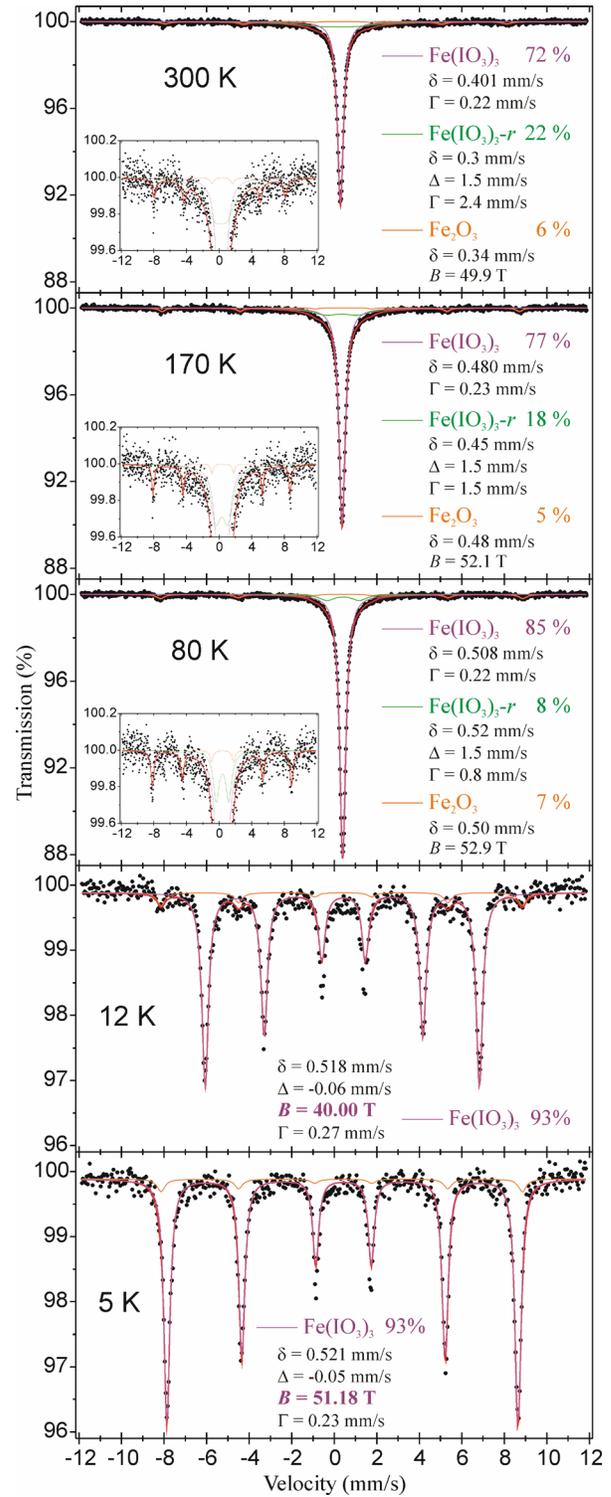

Figure 6. $^{57}$Fe Mössbauer spectra of Fe(IO$_3$)$_3$ at different temperatures. The Mössbauer spectroscopy parameters (the center shift $\delta$ versus room temperature $\alpha$-Fe, the quadrupole splitting $\Delta$, the hyperfine magnetic field $B$, and the absorber linewidth $\Gamma$ within transmission integral approach) are shown together with their respective sub-spectra cross-sectional areas. The sub-spectra marked in orange belongs to traces of Fe$_2$O$_3$. Errors for all values are of the order of unity for the last digit shown.

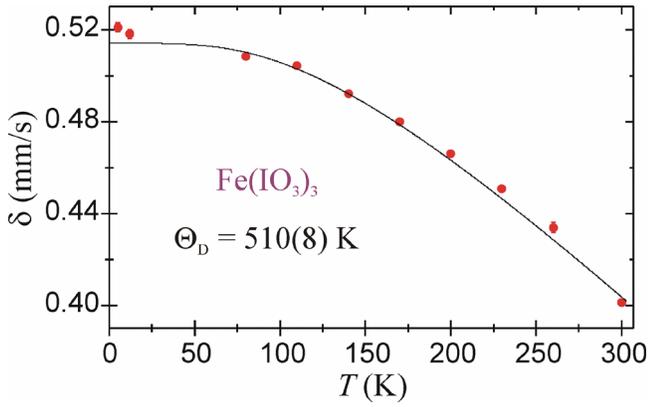

Figure 7. The center shift $\delta$ versus temperature for $^{57}$Fe Mössbauer spectral component of Fe(IO$_3$)$_3$. The solid line represents the best fit to the experimental data according to the standard Debye model for the phonon spectrum and the resulting Debye temperature $\Theta_D$ = 510(8) K.

The temperature dependence of the center shift $\delta$ shown in Figure 7 presents typical second-order Doppler shift variability with temperature (a relativistic effect due to the thermal motion of the absorbing nuclei) and it was considered in terms of the Debye approximation of the lattice vibrations of Fe atoms. By fitting the Mössbauer experimental data it was found that the Debye temperature $\Theta_D$ = 510(8) K of Fe(IO$_3$)$_3$ is slightly higher than typical for inorganic compounds (150–500 K). Such value of the $\Theta_D$ indicates high rigidity of the iron bindings rather typical for metallic systems. The results of the $^{57}$Fe Mössbauer spectra support the thermodynamic signature of the ground state of Fe(IO$_3$)$_3$ obtained from the aforementioned specific heat measurements, that is, the phase transition at $T_N$ = 17 K is completely magnetically driven, not structural.

To understand the Fe(IO$_3$)$_3$ ground state magnetic order, a neutron diffraction experiment was performed on the HB2A diffractometer at Oak Ridge National Laboratory.[28] Data were collected on 5 g of powder at 150 K, 20 K, and 1.6 K (Figure 8a). Comparing the diffraction patterns at different temperatures, there is a buildup of intensity near the (100) Bragg peak at 20 K, and new Bragg peaks appear at $T$ = 1.6 K < $T_N$. The order parameter curve in Figure 8b shows the onset temperature to be 16.2(9) K, in agreement with the transition temperatures seen in magnetic susceptibility and heat capacity, and thus we associate these Bragg peaks with long range magnetic order. The order parameter curve in Figure 8b was fitted to a critical exponent $\beta$ = 0.31(2), which is broadly consistent with three-dimensional magnetic exchange interactions.[49]

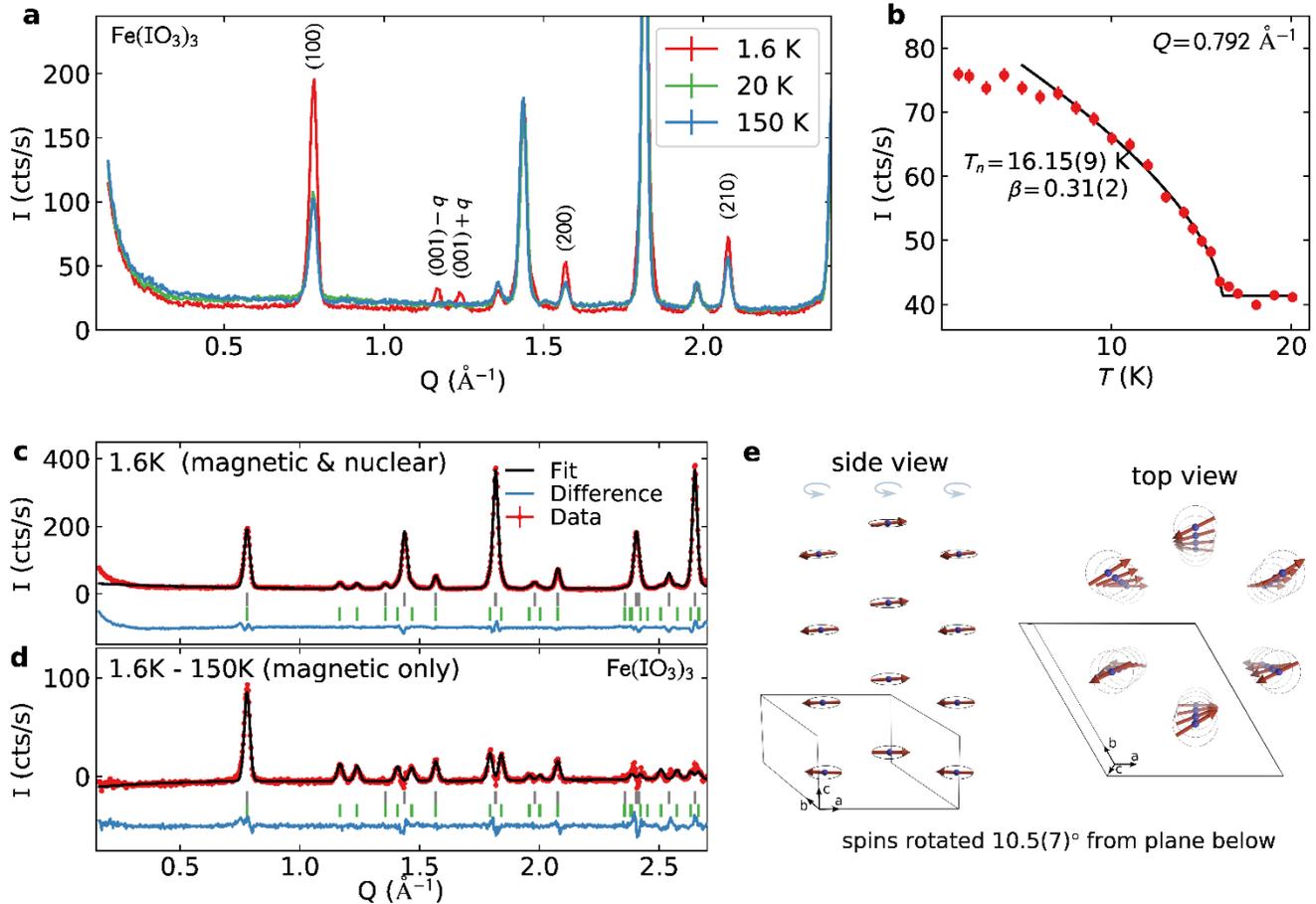

Figure 8. (a) Powder neutron diffraction of Fe(IO$_3$)$_3$ at three different temperatures: 1.6 K, 20 K, and 150 K, showing new Bragg peaks appearing at the lowest temperatures. These peaks are consistent with (0,0,l) order. (b) Magnetic diffraction of the (100) peak as a function of temperature, showing a Neel transition temperature of 16.15(9) K and a critical exponent $\beta = 0.31(2)$, consistent with 3D magnetic exchange interactions.[49] (c) Nuclear and magnetic refinement for Fe(IO$_3$)$_3$ diffraction. (d) Refinement to the temperature-subtracted magnetic only data, showing that the magnetic structure fits the data very well. (e) The fitted spin structure, showing AFM correlations in the *ab* plane and a ferromagnetic incommensurate spiral along the *c* axis.

The magnetic Bragg peaks can be indexed by a single propagation vector $q = (0, 0, 0.0293(1))$, which indicates incommensurate magnetic order along the *c* axis (see the Supporting Information for details). Using BasIreps from the FullProf software package,[50] representational analysis showed six irreducible representations consistent with the propagation vector (Table S6). Subsequent magnetic Rietveld refinement using FullProf (Figure 8c-e) clearly identified the correct magnetic structure to be $\Gamma_2$: AFM correlations in the *a* and *b* directions and an incommensurate ferromagnetic spiral along the *c* axis. Each spin is rotated 10.5(7)° from the spin below. (For plots of the alternative irreducible representations, see Figure S7.) This magnetic structure indicates dominant AFM exchange interactions in the *ab*-plane, and dominant FM exchange interactions along the *c*-axis with an anisotropy causing incommensurability. The FM incommensurate spiral pattern along the *c*-axis repeats at every 18 nm, which is approximately over 34 unit cells (the *c*-axis unit cell dimension of 5.1879(2) Å ~ 0.52 nm). The fitted static magnetic moment is 4.08(2) $\mu_B$, which is 82 % of the full $S = 5/2$ value of 5 $\mu_B$. This indicates some low-lying magnetic excitations and potential entanglement between spins. The discrepancy to the fit at high $Q$ (Figure 8d) comes from subtracted nuclear Bragg peaks, which introduce some noise in the temperature-subtracted data wherever they are. Note that wherever a nuclear Bragg peak is weak or absent, the fit matches the data very well. Thus, the discrepancy near $Q$=2.5 Å$^{-1}$ does not indicate an issue with the fitted model.

Although these diffraction experiments were in zero field, the spiral antiferromagnetic ground state points toward Fe(IO$_3$)$_3$ being a skyrmion material: finite-field skyrmion lattices generally have a spiral AFM zero-field ground state.[51] More specifically, the spiral antiferromagnetism indicates a substantial DM exchange $D \cdot (S_i \times S_j)$ between spins. Due to the lack of inversion symmetry and the three-fold rotation axis along *c* for each Fe ion, Moriya's rules[52] dictate that the *D* vector between Fe spins along the *c* axis is nonzero and oriented along *c*. Such a DM exchange combined with isotropic Heisenberg exchange would tend to produce incommensurate spiral order precisely that seen in Figure 8e. In fact, the minimum energy configuration is easy to compute. Assuming a Hamiltonian:

$$H = JS_i \cdot S_j + D \cdot (S_i \times S_j) \qquad (4)$$

the energy relative to the angle between spins can be minimized and the relative strength of the DM interaction can be estimated from the propagation vector: $\frac{D}{J} = tan(2\pi Q_c) = 0.184 \pm 0.013$. Thus, the DM exchange is estimated to be 18% of the strength of the Heisenberg exchange along the *c* axis. This evidence of appreciable DM exchange and the zero-field spiral AFM order bolsters the hypothesis that the finite-field pocket is indeed a skyrmion phase. The period of the DM-driven incommensurate order is 18 nm, we anticipate that finite-field skyrmions would have small radius, comparable to the helix wavelength. This associated magnetic modulation period of the expected skyrmions in Fe(IO$_3$)$_3$ is within the range of those realized in other host materials.[13, 53-56]

To further gain insight into the chemical bonding and orbital overlapping in Fe(IO$_3$)$_3$, DFT computations were performed and the results are depicted in Figure 9. The density of states (DOS) clearly reveals that Fe-*d* derived states are polarized, which is direct evidence of $3d^5$ Fe$^{3+}$ magnetic cations. The spin of Fe$^{3+}$ also polarizes the O-*p*, I-*s* and I-*p* electrons, forming Fe sub-lattice with AFM couplings. The valence maximum is mostly composed of the Fe-*d*, O-*p*, I-*s* and I-*p* states. The conduction band minimum is mainly derived from the Fe-*d* states. Thus, DFT computations support complex magnetic interactions in this material and manifestation of the O-*p*, I-*s* and I-*p* states participating in the couplings between Fe$^{3+}$ magnetic moments, promoting DM exchange interaction.

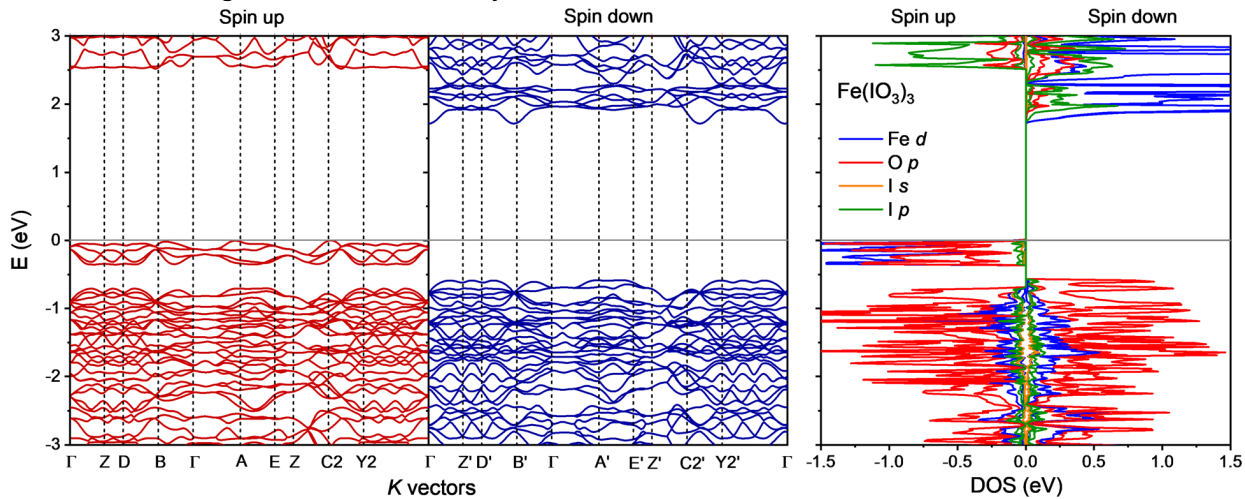

Figure 9. Spin-polarized band structure and density of states (DOS) of $Fe(IO_3)_3$ showing the orbital overlap of Fe-$d$, O-$p$, I-$s$, and I-$p$ states at the valence band maximum.

## CONCLUSION

The design of magnets comprised of materials lacking inversion symmetry with the potential to host topologically non-trivial spin states is an emerging experimental challenge, and opening up a viable avenue to create these materials is a key step forward. The synthesis and growth of $Fe(IO_3)_3$ provides a new experimentally accessible model for preparing both bulk sample and single crystalline forms, potentially supporting transformative materials growth of new magnetic polar compounds. This study focused on the breaking of inversion symmetry in materials by combining metal ion and asymmetric anion building units having stereo-active lone-pair electrons. The results indicate that $Fe(IO_3)_3$ features magnetic high-spin $S$ = 5/2 state and macroscopic polarization. A putative skyrmion phase driven by asymmetric DM exchange emerges at 14 K ≤ $T$ ≤ 16 K and 2.5 T ≤ $\mu_0 H$ ≤ 3.2T, as evidenced by the sizeable DM interaction and the zero-field spiral AFM ground state of $Fe(IO_3)_3$. The increased DM exchange of the $Fe^{3+}$ spins are mediated through the O-$p$, I-$s$ and I-$p$ states. The size of the predicted skyrmions is estimated to be small ∼ 18 nm, akin to the magnetic modulation period of the DM-driven incommensurate order. Stabilizing such compact skyrmions has been a major goal in the filed for spin-based technological applications in high-density memory architectures. Further studies to provide supporting solid evidence for the emergence of skyrmions in $Fe(IO_3)_3$ including Lorentz Transmission Electron Microscopy and Small Angle Neutron Scattering are underway. These results demonstrate how an extended polar lattice of magnetic ions produces exotic physics while providing numerous advantages for accessing new skyrmion host candidates. This approach opens a new frontier in the search for topologically distinct states of matter.

## ASSOCIATED CONTENT

**Supporting Information**.
Crystallographic data obtained by single crystal and synchrotron X-ray diffraction for $Fe(IO_3)_3$; Thermogravimetry, IR spectra for $Fe(IO_3)_3$; Magnetic Rietveld refinement for $Fe(IO_3)_3$. Powder X-ray diffraction spectra for $Ga(IO_3)_3$. This material is available free of charge via the Internet at http://pubs.acs.org.

## AUTHOR INFORMATION

### Corresponding Author
* Thao T. Tran, thao@clemson.edu
Department of Chemistry, Clemson University, Clemson, SC 29634, USA

### Author Contributions
The manuscript was written through contributions of all authors. All authors have given approval to the final version of the manuscript.

### Notes
The authors declare no conflict of interest.

## ACKNOWLEDGMENT
This work was supported by Clemson University, College of Science, Department of Chemistry. EEO acknowledges the COSSAB-GIAR Grant from College of Science, Clemson University. TTT thanks the 2021 Support for Early Exploration and Development (SEED) Grant. Research at Gdansk University of Technology was supported by the National Science Center (Poland) under SONATA-15 grant (no. 2019/35/D/ST5/03769). MJW gratefully acknowledges the Ministry of Science and Higher Education scholarship for young scientists. A portion of this research used resources at the High Flux Isotope Reactor, a DOE Office of Science User Facility operated by the Oak Ridge National Laboratory. Use of the Advanced Photon Source at Argonne National Laboratory was supported by the U. S. Department of Energy, Office of Science, Office of Basic Energy Sciences, under Contract No. DE-AC02-06CH11357. This manuscript has been authored by UT-Batelle, LLC, under contract DE-AC05-00OR22725 with the US Department of Energy (DOE). The US government retains and the publisher, by accepting the article for publication, acknowledges that the US government retains a nonexclusive, paid-up, irrevocable, worldwide license to publish or reproduce the published form of this manuscript, or allow others to do so, for US government purposes. DOE will provide public access to these results of federally sponsored research in accordance with the DOE Public Access Plan (http://energy.gov/downloads/doe-public-access-plan). We thank Dr. C. McMillen for his assistance in single crystal X-ray diffraction measurement.

## ABBREVIATIONS

AFM, Antiferromagnetism; FM, Ferromagnetism; DM, Dzyaloshinskii–Moriya; SOC, Spin-orbit coupling; PXRD, Powder X-ray diffraction; ATR-FTIR, Attenuated total reflection Fourier transform infrared; PPMS, Physical Properties Measurement System; TGA, Thermogravimetric analysis; DSC, Differential scanning calorimetry; DOS, Density of States.

Table of Content

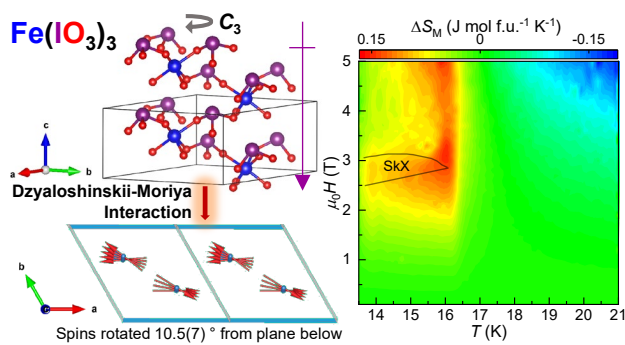

Access to the potential emergence of topologically distinct spin textures (skyrmions) enabled by chemical logic to tune competing magnetic exchange in Fe(IO$_3$)$_3$ is reported. A putative skyrmion phase driven by asymmetric exchange emerges at 14 K ≤ $T$ ≤ 16 K and 2.5 T ≤ $\mu_0H$ ≤ 3.2 T, as evidenced by the appreciable asymmetric interaction and the zero-field spiral antiferromagnetic ground state of Fe(IO$_3$)$_3$ deduced from neutron diffraction experiments. Our results connect stereo-active lone-pair electron effects to enhanced DM interaction, demonstrating a new approach for chemical guidelines in the search for such novel spin states of matter.